\documentclass[aps,prl,pre,twocolumn,superscriptaddress]{revtex4-2}
\usepackage[pdftex]{graphicx}
\usepackage{lipsum}
\usepackage{SIunits}
\usepackage{xspace}
\usepackage{xcolor}
\usepackage{amsmath}

\newcommand{\cnot}{\ensuremath{c_0}\xspace}
\newcommand{\cs}{\ensuremath{c_\text{s}}\xspace}
\newcommand{\cR}{\ensuremath{c_\text{R}}\xspace}
\newcommand{\dc}{\ensuremath{\Delta c}\xspace}
\newcommand{\kh}{\ensuremath{k_\mathrm{H}}\xspace}
\newcommand{\Rc}{\ensuremath{R_\mathrm{c}}\xspace}
\newcommand{\Rp}{\ensuremath{R^\prime}\xspace}

\newcommand{\zmax}{\ensuremath{Z_\mathrm{max}}\xspace}

\newcommand{\thetac}{\ensuremath{\theta_\mathrm{c}}\xspace}

\begin{document}

\title{Universal equation describes the shape of air bubbles trapped in ice}


\author{Virgile Thi\'evenaz}
\email[]{virgile.thievenaz@espci.fr}
\affiliation{PMMH, CNRS, ESPCI Paris, Sorbonne Universit\'e, Universit\'e Paris-Cit\'e, F-75005, Paris, France}
\affiliation{Department of Mechanical Engineering, University of California, Santa Barbara, California 93106, USA}

\author{Jochem G. Meijer}
\affiliation{Physics of Fluids Group and Max Planck Center for Complex Fluid Dynamics, Department of Science and Technology, J.M. Burgers Center for Fluid Dynamics, and MESA+ Institute, University of Twente, Enschede, The Netherlands}
\author{Detlef Lohse}
\affiliation{Physics of Fluids Group and Max Planck Center for Complex Fluid Dynamics, Department of Science and Technology, J.M. Burgers Center for Fluid Dynamics, and MESA+ Institute, University of Twente, Enschede, The Netherlands}
\affiliation{Max Planck Institute for Dynamics and Self-Organization, G\"ottingen, Germany}
\author{Alban Sauret}
\affiliation{Department of Mechanical Engineering, University of California, Santa Barbara, California 93106, USA}
\affiliation{Department of Mechanical Engineering, University of Maryland, College Park, Maryland 20742, USA}
%\email[]{}
%\thanks{}
% \altaffiliation{}


\date{\today}

\begin{abstract}
    Water usually contains dissolved gases, 
    and because freezing is a purifying process these gases must be expelled for ice to form. 
    Bubbles appear at the freezing front and are then trapped in ice, making pores.
    These pores come in a range of sizes from microns to millimeters
    and their shapes are peculiar; 
    never spherical but elongated, and usually fore-aft asymmetric. 
    We show that these remarkable shapes result of a delicate balance 
    between freezing, capillarity, and mass diffusion. 
    A non-linear ordinary differential equation 
    suffices to describe the bubbles,
    which features two non-dimensional numbers representing the supersaturation and the freezing rate,
    and two additional parameters representing simultaneous freezing and nucleation 
    treated as the initial condition.
    Our experiments provide us with a large variety of pictures of bubble shapes. 
    We show that all of these bubbles have their rounded tip well described by an asymptotic regime of
    the differential equation,
    and that most bubbles can have their full shape quantitatively matched by a full solution.
    This method enables the measurement of the freezing conditions of ice samples,
    and the design of freeze-cast porous materials.
    Furthermore, the equation exhibits a bifurcation that explains
    why some bubbles grow indefinitely and make long cylindrical ''ice worms'',
    well known to glaciologists.    
\end{abstract}

%\keywords{}

\maketitle

\begin{figure*}[t]
    \centering
    \includegraphics[width=.95\textwidth]{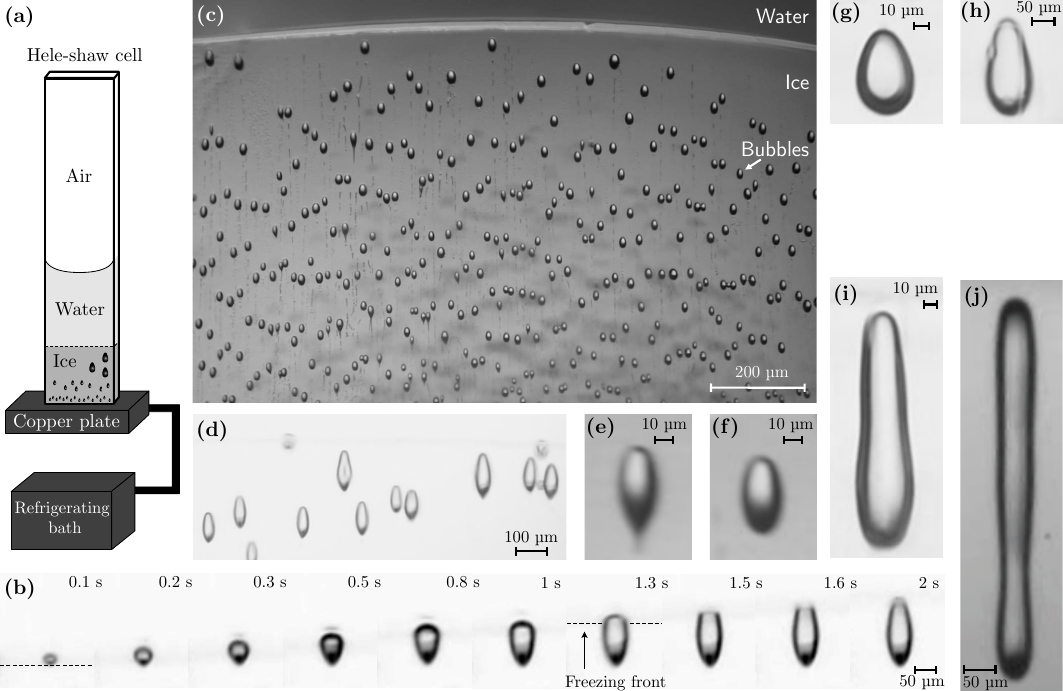}
    \caption{(a) Schematic of the experimental setup.
        (b) Timelapse of the nucleation, entrapment, growth and closure
        of a gas bubble at the ice-water interface, 
        for a freezing rate of \unit{54}\micro\meter\per\second.
        (c) Small pores in ice frozen at 133~\micro\meter\per\second.
        (d) Long pores in ice frozen at 46~\micro\meter\per\second.
        Other pictures describe particular cases:
        (e) $v = 91~\micro\meter\per\second$, with nucleation between ice grains;
        (f) $v = 90~\micro\meter\per\second$,
        with nucleation at the freezing front;
        (g) $v = 21~\micro\meter\per\second$, 
        with nucleation $35\micro\meter$ away from the freezing front;
        (h) $v = 20~\micro\meter\per\second$, 
        with nucleation $100\micro\meter$ away from the freezing front;
        (i) short ice worm for $v = 13~\micro\meter\per\second$;
        (j) Worm bubble from second experimental setup (see Materials and Methods).
    }
    \label{fig:shapes}
\end{figure*}

%%%%%%%%%%%%%%% There are air bubbles in ice
Ice frozen from water containing dissolved air is usually not clear but opaque,
because it includes many bubbles~\cite{warren2019optical}.
This is commonly observable in ice cubes from a freezer.
These bubbles have peculiar shapes, never spherical but elongated (Fig.~\ref{fig:shapes}).
Some even reach lengths of several centimeters~\cite{murakami2002}
-- they are named ''ice worms''~\cite{chalmers1959water}
or ''worm bubbles''~\cite{swinzow1966ice}.
Gases are soluble in liquid water but not in ice,
so that when water freezes the dissolved gases
are expelled and concentrate in the liquid~\cite{wei2003}.
Bubbles eventually nucleate near the freezing front and are captured by ice,
while at the same time they keep growing by diffusion of the gas.
Ice thus formed is porous.
Usually, one speaks of bubbles in water and pores in ice.

%%%%%%%%%%%%%%% Glaciology
% One usually speaks of bubbles in a liquid and of pores in a solid.
In the natural environment, porous ice is the rule rather than the exception.
Hailstones~\cite{bari1974} and lake ice~\cite{swinzow1966ice,gow1977growth} contain pores
made out of the dissolved gas.
Glacier ice is also porous but is made out of compacted snow
and not frozen gas-laden water~\cite{wengrove2023melting}.
%%%%%%%%%%%%%%% Winter embolism in trees
% 
In winter, sap freezes inside plants and bubbles form;
after the thaw these bubbles may prevent the flow of sap (winter embolism)~\cite{charra2023xylem}.
%%%%%%%%%%%%%%% Freezing as preservative
Generally, freeze-thaw cycles can dramatically affect the stability 
of complex media, like food~\cite{ghosh2008factors},
and the survival of living organism~\cite{korber1988phenomena,kletetschka2015}.

In addition to water, gases are soluble in a large variety of liquids,
including metals~\cite{shapovalov2004}, silica~\cite{yokokawa1986gas}
and sapphire~\cite{bunoiu2010,ghezal2012};
freezing such gas solutions yields porous materials.
Porosity is usually a defect of which to get rid~\cite{gupta1992,bianchi1997}.
% 
%%%%%%%%%%%%%%% Freeze-cast materials (Sylvain Deville, Lohse, etc\ldots)
However, for certain applications porosity is desired and therefore the
size and shape of the pores must be controlled~\cite{nakajima2001fabrication}.
More generally, the freezing of solutions of gases or other solutes
makes various freeze-cast materials~\cite{deville2006,deville2007ice,deville2008freeze},
some of which are biocompatible~\cite{deville2010freeze}.
During the freezing of a suspension or of an emulsion,
the dispersed particles may or may not be engulfed in ice, 
depending on the freezing rate~\cite{uhlmann1964}.
Their engulfment deforms the freezing front~\cite{asthana1993},
according to their thermal properties~\cite{tyagi2020}.
The particles themselves may deform when they are captured;
for example, oil droplets in an emulsion make pointy oil drops
in ice~\cite{tyagi2022solute,meijer2023thin}.

%%%%%%%%%%%%%%% Modeling: Wei 
Several attempts at describing the growth and entrapment
of gas bubbles have been made,
either using scaling laws ~\cite{shao2023growth},
or taking into account the numerous mechanisms at play 
(heat transfer, phase change, capillarity, mass diffusion, nucleation)
~\cite{wei2000,wei2003,wei2017}.
Freezing and capillarity make a challenging combination.
For example, a sessile drop freezing will grow a 
tip~\cite{anderson1996,snoeijer2012,marin2014,seguy2023role}.
Also, ice is actually not perfectly hydrophilic, so that water
may retract on ice instead of spreading~\cite{knight1967,thievenaz2020a,huerre2021solidification}.

%%%%%%%%%%%%%%% This paper
In this paper, we investigate the shape of the pores formed during the freezing 
of ice, and how it is set by the growth history of bubbles.
We show that under certain assumptions, this problem reduces to a single non-linear
ordinary differential equation, which we study analytically.
Two asymptotic regimes are found, 
one corresponding to fast freezing and the other to the closing of the bubble.
Under a certain freezing velocity the system undergoes a bifurcation,
after which bubbles do not close any more, thus explaining how worm bubbles appear and what their equilibrium radius is.
Our equation can also be solved numerically,
and its solutions matched to the shapes of pores
obtained experimentally by freezing deionized water at various freezing rates.
In most cases a quantitative agreement is found 
between the solution and the experiments.

% \lipsum

%%%%%%%%%%%%%%%%%%%%%%%%%%%%%%%%%%%%%%%%%%%%%%%%%%%%%%%%%%%%%%%%%%%%%%%%%%%%%%%%
%%%%%%%%%%%%%%%%%%%%%%%%%% EXPERIMENTS %%%%%%%%%%%%%%%%%%%%%%%%%%%%%%%%%%%%%%%%%
%%%%%%%%%%%%%%%%%%%%%%%%%%%%%%%%%%%%%%%%%%%%%%%%%%%%%%%%%%%%%%%%%%%%%%%%%%%%%%%%
\section*{Experiments}

Much work having already been dedicated to describing the pores in ice 
\cite{bari1974,wei2000,murakami2002,wei2003,wei2004,wei2017,shao2023growth},
the main purpose of our experiments is to provide a basis of 
comparison with our model,
as well as a qualitative description for the reader to
better understand the problem and the assumptions of the model.

We study the freezing of water, in which gases are naturally dissolved,
in a Hele-Shaw cell for freezing rates varying between 12~\micro\meter\per\second\ and 263~\micro\meter\per\second\ 
(Fig.~\ref{fig:shapes}a, and Materials and Methods).
Figure~\ref{fig:shapes}(b) reports the formation of a pore at a freezing rate
of $54~\micro\meter\per\second$.
First, a bubble nucleates at the freezing front.
Then, this bubble grows by diffusion of dissolved gas,
while the freezing front keeps advancing.
The bubble expands radially up to a maximum after which it shrinks.
Eventually, the freezing front passes by and the pore closes; its final shape is set.
For the case shown in Figure~\ref{fig:shapes}(b), the whole process takes about 
2~s, and the final bubble is 154 \micro\meter-long
and 64 \micro\meter-wide.

The pores in ice are never spherical,
but elongated in the direction of freezing.
Their number also varies with the freezing rate.
At a high freezing rate (Fig.~\ref{fig:shapes}c, 133~\micro\meter\per\second\ on average),
many small pores, slightly elongated, are formed.
Conversely, at a slower freezing rate (Fig.~\ref{fig:shapes}d, 46~\micro\meter\per\second), 
fewer pores are formed, and they are bigger and longer.

The place of nucleation of the bubble and the time elapsed before it is trapped
may also influence greatly its final shape.
For fast freezing (about 90~\micro\meter\per\second) the bubble may
(Fig.~\ref{fig:shapes}e) or may not (Fig.~\ref{fig:shapes}f) show a ''tail''.  
This likely corresponds to whether or not the bubble was trapped by a single ice crystal 
or at the junction of two crystals, \textit{i.e.} at a grain boundary.
In the latter case, the initial growth of the bubble is restrained between two crystals,
and this gives it the tail.

Although nucleation usually occurs at the freezing front,
we observe some cases in which the bubble nucleates ahead of the front,
probably on some tiny impurity 
(see Fig.~\ref{fig:shapes}g-h and the corresponding videos) \cite{meijer2024enhanced}.
For slow freezing, 
the distance between the nucleation point and the freezing front matters greatly because 
it sets the time span during which the bubble may grow before being trapped.
For instance, the pores shown in Figs.~\ref{fig:shapes}(g) and (h) grew 
and froze under the same freezing rate of 20~\micro\meter\per\second, 
however their shapes are different.
The former is 53 \micro\meter-long and 36 \micro\meter-wide,
whereas the latter is 200 \micro\meter-long and 96 \micro\meter-wide;
its aspect ratio is larger.
The difference is that the corresponding bubbles
nucleated 35 \micro\meter\ away from the freezing front (Fig.~\ref{fig:shapes}g),
and 100 \micro\meter\ away from it (Fig.~\ref{fig:shapes}h).
The latter had about 3~s more to grow freely,
eventually yielding a different shape, not only a different size.
This observation reveals the strong dependence of the bubble shape 
on the initial bubble size.

The nucleation of gas bubbles in water is quite complex 
because of the chemistry it involves~\cite{lubetkin1995fundamentals}.
In the following, we shall focus on the growth of the bubble at the freezing front,
after the nucleation and the trapping.
As we shall show, the growth can be well described by a single
ordinary differential equation.
Nucleation and entrapment will appear as initial conditions.

Our first experimental setup, from which the pictures of Figures \ref{fig:shapes}(a-i) 
and most of our experimental data were obtained, 
could not maintain freezing rates slower than 12~\micro\meter\per\second\ with sufficient stability.
The fluctuations of the freezing rate in this system are indeed
of the order of a few \micro\meter\per\second.
We studied worm bubbles using another experimental setup, 
described in Material and Methods.
This second system enabled us to vary the freezing rates between
4~\micro\meter\per\second\ and 17~\micro\meter\per\second\ 
with a stability of the order of a few tenths of \micro\meter\per\second. 
We used this setup to generate worm bubbles 
(Fig. \ref{fig:shapes}j and the later Fig. \ref{fig:worms}).

%%%%%%%%%%%%%%%%%%%%%%%%%%%%%%%%%%%%%%%%%%%%%%%%%%%%%%%%%%%%%%%%%%%%%%%%%%%%%%%%
%%%%%%%%%%%%%%%%%%%% THE FROZEN BUBBLE EQUATION %%%%%%%%%%%%%%%%%%%%%%%%%%%%%%%%
%%%%%%%%%%%%%%%%%%%%%%%%%%%%%%%%%%%%%%%%%%%%%%%%%%%%%%%%%%%%%%%%%%%%%%%%%%%%%%%%
\section*{The Frozen Bubble Equation}

\begin{figure}[t!]
    \centering
    \includegraphics[width=\linewidth]{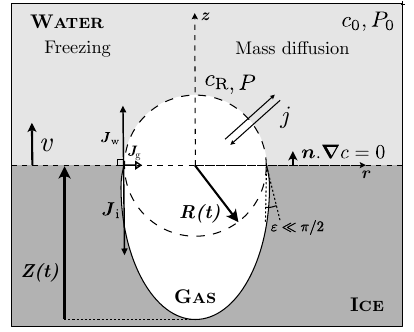}
    \caption{Schematic of the model.
    The upper part of the gas is the bubble, in contact with water;
    the lower part is the pore, in contact with ice.
    On the left thermal effects are shown, on the right mass diffusion effects.
    }
    \label{fig:model}
\end{figure}

%%%%%%%%%%%%%%%%%%%%%%%%%% DERIVATION %%%%%%%%%%%%%%%%%%%%%%%%%%%%%%%%%%%%%%%%%%
% \subsection*{Derivation}

%%%%%%%%%%%%%%% General Idea
Our model is based on the conservation of the mass of gas.
For simplicity, we treat air as a simple gas 
and average its properties over that of nitrogen and oxygen
(see Material and Methods).
During its growth, the bubble can be separated into two parts:
a lower part, the pore, that is trapped in the ice,
and an upper part, the bubble \emph{per se},
that is in contact with the liquid water (see Fig.~\ref{fig:model}).
As water freezes, gas is virtually transferred from the bubble to the pore.
In the meantime, gas may be exchanged between the bubble 
and the surrounding water.
The direction and magnitude of this mass transfer depends on the Laplace pressure,
hence on the curvature of the bubble.

%%%%%%%%%%%%%%% Thermal fluxes and the Stefan condition
The shape of the ice-water interface results of the balance between 
latent heat and the heat fluxes on either side.
Since the heat conductivity of gas is negligible compared to that of water and ice,
the heat flux towards the bubble $\boldsymbol{J}_\mathrm{g}$ must be negligible as well,
hence the streamlines of the heat flux must be tangent to the bubble.
Being the isotherm corresponding to the freezing temperature, 
the freezing front should be orthogonal to the streamlines and therefore orthogonal to the bubble.
Therefore, we will assume that the contact angle of the bubble on the ice is close to $\pi/2$,
as observed for instance in the freezing of sessile drops \cite{marin2014,seguy2023role}.
The difference between the actual contact angle and $\pi/2$ (denoted $\varepsilon$ in Fig. \ref{fig:model})
will be neglected; this assumption is supported \emph{a posteriori} by our measurements.
Since the bubble is much smaller than the capillary length (2.7 mm for water),
the liquid-gas interface has a uniform curvature; it is a spherical cap.
In summary, heat transfer and capillarity impose that the upper part of the bubble be a hemisphere
of radius $R$.
Let $R$ be the radius of that hemisphere, and also the radius of the contact line between the bubble and 
the freezing front.
The rest of the model consists in writing an evolution equation for $R(t)$ and 
then in constructing the shape of the pore by translation of the contact line of radius $R(t)$ at velocity $v$.

In order to simplify the calculations, 
we assume that the gas concentration far away from the bubble, $c(r\to\infty) = \cnot$, is constant.
In practice, gas is released into the liquid at the freezing front,
and that creates a concentration profile in the z-direction with a strong 
gradient near the front.
More specifically, $c(z)/\cnot \sim \exp \left( -z\,v/D \right)$ \cite{bari1974}, with $D$ the gas diffusivity.
Meijer \textit{et al.} have recently estimated the gradient on similar experiments \cite{meijer2024enhanced};
they find that it is of order 1~$(\mathrm{mg}/\mathrm{L})/\mathrm{mm}$.
For the gas concentrations considered here, it means that the concentration profile
near the front far from the bubble smoothes out over a length scale $D/v$, which is of order of 10 \micro\meter, about the size of the bubble.
Furthermore, for not too slow $v$, gas accumulates at the front during the freezing process, so that for large pores
frozen over a long time the surrounding gas concentration may increase in time.
The physical meaning of \cnot is therefore closer to an effective concentration,
which averages out the concentration gradients created
by phenomena others than the growth of the current bubble.

%%%%%%%%%%%%%%% Diffusion Around a Hemisphere set on a Plane – Mass Flux
The bubble grows out of the gas-saturated water by mass transfer.
The corresponding mass flux can be expressed in a closed form by solving
the diffusion equation in the half-space bounded by a plane upon which
sits a hemisphere.
Under the hypothesis that the boundary condition on the plane is that
of zero normal flux,
the problem reduces to the rotation-invariant problem of diffusion around sphere.
In spherical coordinates this problem is written
\begin{align}
%     &D \Delta c = \frac{\partial c}{\partial t}\\
    & \frac{D}{r^2} \frac{\partial}{\partial r} \left( r^2 \frac{\partial c}{\partial r}\right)
    = \frac{\partial c}{\partial t},\\
    &c(t=0, r) = \cnot, \\
    &c(t, r=R) = \cR,   \\ 
    &c(t, r\to\infty) = \cnot, 
    \label{eq:diffusion}
\end{align}
where \cnot and \cR denote the boundary conditions at infinity and at the bubble interface, respectively,
and where $D = 2\cdot10^{-9}\mathrm{m}^2/\mathrm{s}$ 
is the diffusion coefficient of air in water~\cite{lide2012}.
This problem was solved by Epstein and Plesset~\cite{epstein1950},
and they obtained a closed form for the surface density of mass flux
through the gas-water interface:
\begin{equation}
    j = D (\cnot-\cR) \left( \frac{1}{R} + \frac{1}{\sqrt{\pi D t}} \right).
    \label{eq:flux}
\end{equation}
The origin of times in Eq. (\ref{eq:flux}) corresponds to when diffusion starts, 
that is when the bubble nucleates.
The solution of the Epstein and Plesset equation is applicable for a bubble in an infinite space.
In our experiments bubbles are ''confined'' for easier visualization,
but the gap between the glass slides is at least ten times bigger than the bubbles, so the Epstein and Plesset equation is still approximately applicable.

%%%%%%%%%%%%%%% Laplace Pressure and Solubility
Far from the bubble, water is at ambient pressure $P_0$, 
and gas is in excess by a quantity $\dc>0$ with respect to the solubility:
$\cnot = \cs(P_0) + \dc$.
Hydrostatic pressure is negligible, given the height of 
the water column (a few centimeters) above the bubble.
The solubility of gases in water \cs depends linearly on pressure through Henry's law:
$\cs(P) = \cs(P_0) + \kh (P-P_0)$.
The constant \kh for air is $2.95 \times 10^{-5} \mathrm{kg}.\mathrm{Pa}^{-1}.\mathrm{m}^{-3}$,
as calculated from the solubility of nitrogen and oxygen~\cite{lide2012}.
At the gas-water interface, the Laplace pressure leads to an increase of 
the solubility:
$\cR = \cs(P_0) + 2\gamma \kh / R$.
$\gamma = 75\ \mathrm{mN}/\mathrm{m}$ is the surface tension of water
at $0^\circ\mathrm{C}$.
Therefore, the concentration gap between the interface and the surrounding water is
\begin{equation}
    \cnot - \cR = \dc - 2\kh\gamma/R.
    \label{eq:concentration}
\end{equation}

%%%%%%%%%%%%%%% Mass Balance 
We can now write the conservation of the mass of gas in the upper part of the bubble:
\begin{equation}
    \frac{\mathrm d}{\mathrm d t} \left(\frac{2\pi}{3}\rho R^3\right)
    = 2\pi R^2 j - \pi R^2\rho v.
    \label{eq:mass_balance}
\end{equation}
$\rho = 1.2\: \mathrm{kg}/\mathrm{m}^3$ is the density of air.
The density mismatch between water and ice is accounted for by the freezing rate $v$.
Substituting the expression of the flux density $j$ (Eq.~\ref{eq:flux} and \ref{eq:concentration})
into the mass balance, and introducing $\Rc = 2\gamma\kh/\dc$ we obtain
\begin{equation}
    \frac{\mathrm d R}{\mathrm d t} = 
        \frac{D \dc}{\rho\Rc} \left(1 - \frac{\Rc}{R}\right) 
        \left( \frac{\Rc}{R} + \frac{\Rc}{\sqrt{\pi D t}} \right)
        - \frac{v}{2}.
    \label{eq:dRdt}
\end{equation}
Let us define $Z(t)$ so that 
\begin{equation}
    \frac{\mathrm d Z}{\mathrm d t} = v,
    \label{eq:dZdt}
\end{equation}
then $\left[Z(t), R(t)\right]$ is a parametric curve
that describes the shape of the pore.

A direct consequence of Eq.~\ref{eq:dRdt} is that the bubble can only grow
if $R>\Rc$.
This observation reveals the physical meaning of \Rc: it is the critical radius a bubble
must have to be stable against dissolution under Laplace pressure.
Therefore, even if the water around the bubble is supersaturated with gas, 
the bubble may still dissolve if it is too small.
The condition $R>\Rc$ is nevertheless not sufficient to maintain stability,
because freezing contributes negatively to $\frac{\mathrm d R}{\mathrm d t}$.

%%%%%%%%%%%%%%% Non-dimensional form
In the rest of the paper, we shall restrict ourselves to the case of constant freezing rate
($\frac{\mathrm d v}{\mathrm d t} = 0$).
Not only does this simplify greatly the analysis,
it also corresponds to our experimental situation.
Therefore, Eq.~(\ref{eq:dZdt}) becomes trivial and its solution $Z = vt$ can
be combined with Eq.~(\ref{eq:dRdt}) to reduce the problem to a single
non-linear ordinary differential equation.
We write this equation in non-dimensional form by taking \Rc as unit length 
and $\Rc^2/D$ as unit time:
\begin{equation}
    \frac{\mathrm d R}{\mathrm d Z} =
    \frac{\delta}{\zeta} \left( 1 - \frac{1}{R} \right) \left( \frac{1}{R} + \sqrt\frac{\zeta}{\pi Z}\right)
    - \frac{1}{2}.
    \label{eq:frozen}
\end{equation}
We refer to Eq.~(\ref{eq:frozen}) as the Frozen Bubble Equation.
It has two non-dimensional parameters:
$\delta = \dc/\rho$ is the non-dimensional supersaturation,
and $\zeta = v\Rc/D$ is the ratio of the characteristic times of freezing and mass diffusion.
In the following, we denote $\Rp = \frac{\mathrm{d} R}{\mathrm{d} Z}$.

\begin{figure}[h]
    \centering
    \includegraphics[width=\linewidth]{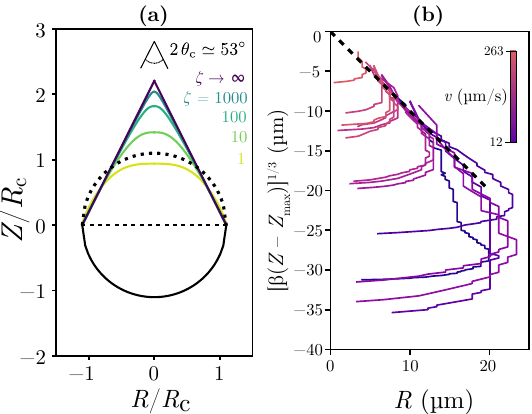}
    \caption{(a) Solutions of Eq.~(\ref{eq:frozen}) in the limit $\zeta \gg \delta$,
    for $\delta = 1$ and various values of $\zeta$.
    The case $\zeta \to \infty$ is the analytical solution.
%     (b) Profiles of pores frozen at different freezing rates in our experimental range.
    (b) Rescaling of the shape of the pores near the tip according to Eq.~(\ref{eq:Rasy}).
    The prefactor $\beta$ is fitted for each profile.
    The dashed line has slope $-1$.
    }
    \label{fig:asympt}
\end{figure}

%%%%%%%%%%%%%%%%%%%%%%%%%%%%%%%%%%%%%%%%%%%%%%%%%%%%%%%%%%%%%%%%%%%%%%%%%%%%%%%%
%%%%%%%%%%%%%%%%%%%%%%%%%%% ANALYSIS %%%%%%%%%%%%%%%%%%%%%%%%%%%%%%%%%%%%%%%%%%%
%%%%%%%%%%%%%%%%%%%%%%%%%%%%%%%%%%%%%%%%%%%%%%%%%%%%%%%%%%%%%%%%%%%%%%%%%%%%%%%%
\section*{Analysis}

%%%%%%%%%%%%%%%%%%%%%%%%%% FAST FREEZING ζ -> ∞ %%%%%%%%%%%%%%%%%%%%%%%%%%%%%%%%
\subsection*{The fast freezing regime ($\zeta \gg \delta$)}
 
Before turning to the general analysis of the Frozen Bubble Equation,
we describe the specific case in which mass diffusion is negligible
compared to freezing.
Taking $\zeta \to \infty$, Eq.~(\ref{eq:frozen}) reduces to $\Rp = -1/2$.
The upper part of the pore then has the shape of a cone of angle
$\thetac = \arctan(1/2) \simeq 26.5^{\circ}$~(Fig.~\ref{fig:asympt}a).
For finite $\zeta \gg \delta$, we solve the Frozen Bubble Equation numerically 
to check that the shapes of the pores indeed converge towards a conical tip.

Taking the limit $\zeta \to \infty$ amounts to canceling the possibility of gas transfer
between the bubble and the liquid, meaning the two substances are immiscible.
It is interesting to note that in this limit case, the volume of the cone must be that
of a hemisphere of radius $R$, so the mass of gas is conserved.
Satisfying this condition requires that the height of the cone be equal to its maximal
diameter, a condition strictly equivalent to $\thetac = \arctan(1/2)$.
Therefore, in the absence of heat and mass transfer between the bubble and the liquid,
the final pore has the shape of a cone on top of a hemisphere,
and its aspect ratio is $3/2$.
This shape is quite similar to that of pointy oil drops 
in frozen emulsions~\cite{tyagi2022solute,meijer2023thin}, 
although in that case there is no reason to assume a $90^\circ$ contact angle
between oil and the freezing front.

%%%%%%%%%%%%%%%%%%%%%%%%%%%%%% R -> 0 %%%%%%%%%%%%%%%%%%%%%%%%%%%%%%%%%%%%%%%%%%
\subsection*{The limit $R \to 0$ and the tip of the pore}

Another interesting regime is the limit $R \to 0$.
It corresponds to the closing of the pore.
Taking $R \ll 1$ in the Frozen Bubble Equation (\ref{eq:frozen}), we obtain
\begin{equation}
    \frac{\mathrm d R}{\mathrm d Z} = -\frac{\delta}{\zeta}\frac{1}{R^2},
    \label{eq:Rto0}
\end{equation}
which is readily integrated near the closing point of the bubble $R(\zmax)=0$.
Near the tip the pore shape should therefore follow :
\begin{equation}
    R(Z) = \left( \frac{3\delta}{\zeta} \left(\zmax-Z\right) \right)^{1/3}.
    \label{eq:Rasy}
\end{equation}

This asymptotic regime is indeed observed in all our experiments,
without exceptions.
Figure~\ref{fig:asympt}(b) shows a selection of pore profiles extracted from experiments,
representative of the whole range of freezing rates,
rescaled according to Eq. (\ref{eq:Rasy}).
For all pores we obtain a good agreement with Eq.~(\ref{eq:Rasy}),
which in dimensional form is written
$R = \left[ \beta \left(Z_\mathrm{max}-Z\right)\right]^{1/3}$
with $\beta = 3\delta\Rc^2/\zeta$. 
It is notable that $\beta$ does not depend on \dc.
The closing of the pore proceeds regardless of the supersaturation,
only driven by the Laplace pressure.
Below (see Shape Matching), we overlay the cubic shape described by 
Eq. (\ref{eq:Rasy}) onto the profiles.

%%%%%%%%%%%%%%%%%%%%%%%%%%% ALTERNATIVE FORM %%%%%%%%%%%%%%%%%%%%%%%%%%%%%%%%%%%
\subsection*{Alternative form of the Frozen Bubble Equation}

The set of values for parameters $\delta$ and $\zeta$ supplemented with
an initial condition $R_0 = R(Z_0)$ makes a unique solution of Eq.~(\ref{eq:frozen}).
To compare solutions to experimental pores, it is more practical to introduce the initial 
slope $\Rp(Z_0) = \Rp_0$, and to express the ratio $\delta/\zeta$ as a function
of the four parameters $R_0$, $Z_0$, $\Rp_0$ and $\zeta$.
Eq.~(\ref{eq:frozen}) can thus be recast in an elegantly symmetric form:
\begin{equation}
%     \frac{\frac{\mathrm d R}{\mathrm d Z} + \frac{1}{2}}{\Rp_0 + \frac{1}{2}}
    \frac{\Rp(Z) + \frac{1}{2}}{\Rp_0 + \frac{1}{2}}
    =
    \frac{ 1-\frac{1}{R}}{ 1-\frac{1}{R_0}}
    \times
    \frac{\frac{1}{R} + \sqrt\frac{\zeta}{ \pi Z }}{\frac{1}{R_0} + \sqrt\frac{\zeta}{ \pi Z_0 }}.
    \label{eq:frozen_alt}
\end{equation}
$R_0$, $Z_0$ and $\Rp_0$ are geometrical quantities that can be measured
experimentally.

%%%%%%%%%%%%%%%%%%%%%%%%%%%%% BIFURCATION %%%%%%%%%%%%%%%%%%%%%%%%%%%%%%%%%%%%%%
\subsection*{The bifurcation at $\delta = 2\zeta$ and bubbles that never close}
\begin{figure}[t!]
    \centering
    \includegraphics[width=\linewidth]{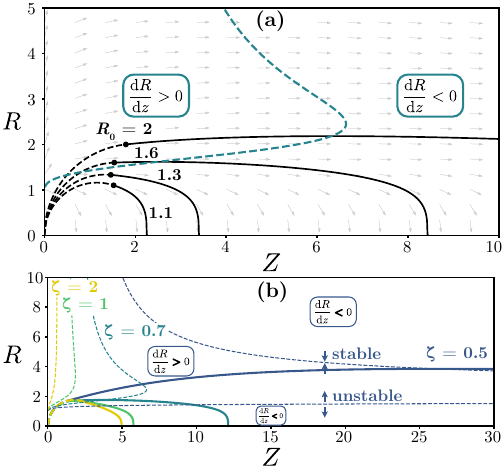}
    \caption{(a) Vector field associated with Eq.~(\ref{eq:frozen}) 
        for $\delta=1$ and $\zeta=0.7$.
        It draws a phase space with domains
        where the bubble grows ($\Rp>0$, left) and shrinks ($\Rp<0$, right).
        The blue dashed line is the separatrix between the domains.
        The black points represent different initial conditions continued to
        the right into solutions (solid lines).
        The dashed curves represent a spherical shape that matches the initial
        condition.
        (b) Evolution of the phase space and the separatrix with decreasing $\zeta$
        for constant $\delta=1$.
        }
    \label{fig:vector}
\end{figure}

%%%%%%%%%%%%%%% Vector field of the frozen bubble shape equation
The general analysis of the solutions of the non-linear Frozen Bubble Equation
can be performed using geometrical techniques.
In the $(Z, R)$-plane, Eq.~(\ref{eq:frozen}) defines a vector field;
at point $(Z,R)$ the vector orientation is $\Rp(Z)$~\cite{arnold1992}.
Starting with the initial condition $R(Z_0) = R_0$, 
the solution is then the curve that passes through $(Z_0, R_0)$
and is everywhere tangent to a vector of the field.
An example is shown in Fig.~(\ref{fig:vector}a) for $\delta=1$ and $\zeta=0.7$.
The dashed blue curve is the separatrix between two domains.
On the left, $\Rp>0$ so the bubble grows;
on the right, $\Rp<0$ so the bubble shrinks.
Starting from different initial conditions, 
the bubble will grow for some time before it shrinks,
or it will shrink right from the start without growing.

Taking $\Rp=0$ in Eq.~(\ref{eq:frozen}), we find the equation of the separatrix:
\begin{equation}
    Z = \frac{\zeta}{\pi}
    \left(\frac{R\left(R-1\right)}{ \left(\frac{\zeta}{2\delta}\right)R^2 - R + 1}\right)^2.
    \label{eq:separatrix}
\end{equation}
The relevant properties of the phase space associated to 
Eq.~(\ref{eq:frozen}) are the number and the shape of the
domains in which the slope \Rp keeps a constant sign.
For $\delta < 2 \zeta$, the separatrix is bounded on the Z-axis and the domain to its right
corresponds to $\Rp<0$ (Fig.~\ref{fig:vector}b).
Therefore, after some possible growth depending on the initial condition, 
the bubble must shrink and close.
At $\delta = 2 \zeta$ there is a bifurcation: the extent of the separatrix on the Z-axis diverges.
For $\delta > 2 \zeta$, the function $Z(R)$ defined by Eq.~(\ref{eq:separatrix}) has two real poles at 
\begin{equation}
%     R_{\pm} = \frac{\delta}{\zeta} \left(1 \pm \sqrt{1 - \frac{2\zeta}{\delta}}\right),
    R_{\pm} = \frac{\delta}{\zeta} \pm \sqrt{\frac{\delta}{\zeta} - 2},
    \label{eq:poles}
\end{equation}
corresponding to two disjoint branches of the separatrix.
Above the lower branch $\Rp > 0$ and below it $\Rp<0$, therefore it is unstable.
However, the upper branch is stable.
If the initial condition is above the lower branch ($R_0 > R_-$), the bubble will grow
until it is captured by the upper branch, and it will never close,
making a worm bubble of equilibrium radius $R_+$ (Fig.~\ref{fig:vector}b).

In reality, worm bubbles have a finite length.
Within the model this would be possible if $\delta$ or $\zeta$ 
would fluctuate so much that the system would switch domains
in Fig.~\ref{fig:vector}b.
Inverting Eq.~(\ref{eq:poles}) gives the typical size of the fluctuation
of $\delta/\zeta$ required to close the worm bubble of a certain radius.
It is notable that the ratio $\delta/\zeta$ scales like $\dc^2$. 
Minute variations of concentration may thus affect greatly
the shape of the bubbles.
Such variations could come from fluctuations of the freezing rate.
It could also come from fluctuations of the gas concentration due to the nucleation
and growth of other neighboring bubbles.
Furthermore, pressure variations in the liquid during the growth are known to
modulate the radius of worm bubbles~\cite{murakami2002}.
Such pressure variations could be taken into account in our model
by modifiying Eq.~\ref{eq:concentration}.

%%%%%%%%%%%%%%% Bifurcation for δ = 2ζ

%%%%%%%%%%%%%%%%%%%%%%%%%%%%%%%%%%%%%%%%%%%%%%%%%%%%%%%%%%%%%%%%%%%%%%%%%%%%%%%%
%%%%%%%%%%%%%%%%%%%%% SHAPE MATCHING %%%%%%%%%%%%%%%%%%%%%%%%%%%%%%%%%%%%%%%%%%%
%%%%%%%%%%%%%%%%%%%%%%%%%%%%%%%%%%%%%%%%%%%%%%%%%%%%%%%%%%%%%%%%%%%%%%%%%%%%%%%%

\begin{figure}[p!]
    \centering
    \includegraphics[width=.90\linewidth]{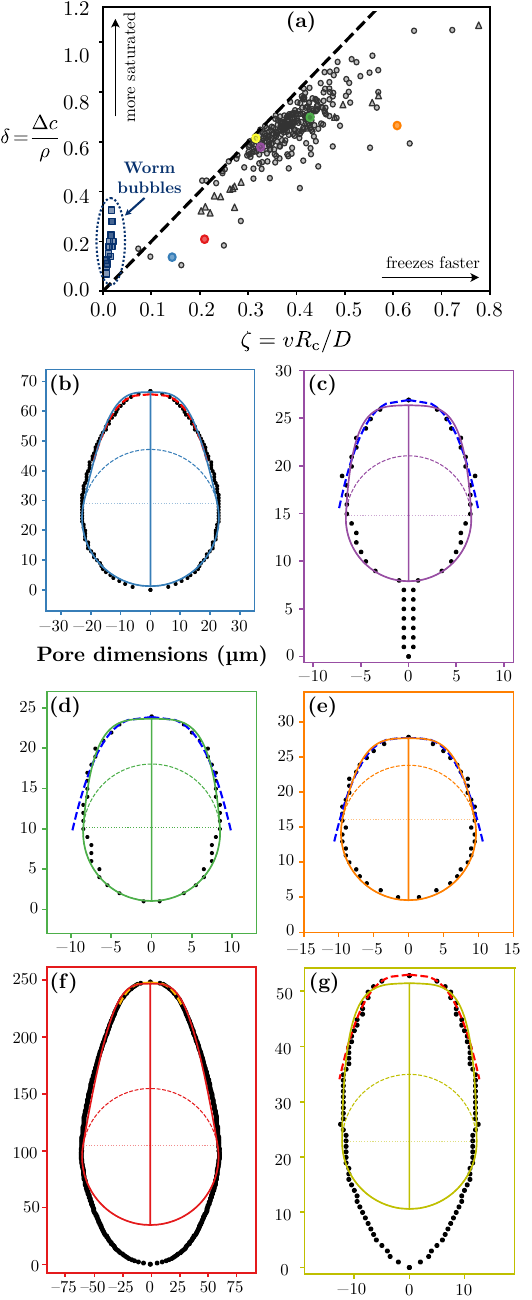}
    \caption{(a) Phase diagram of the experiments.
        Each pore is represented by a point in the ($\zeta$,$\delta$) plane,
        with values obtained by matching a solution of the 
        Frozen Bubble Equation (Eq.~\ref{eq:frozen_alt}) to its shape.
        Circles: first experimental system; triangles and squares: second system.
        Dark blue squares represent worm bubbles.
        The dashed line represents the critical value $\delta = 2\zeta$
        above which the bifurcation to worm bubbles is predicted to occur.
        (b-g) Examples of shape matching, spanning the range of $\delta$ and $\zeta$. 
        The dashed circle represents the spherical initial condition.
        The thin dashed line inside the circle represents the initial condition $R_0, Z_0$ above which
        Eq.~(\ref{eq:frozen_alt}) is solved;
        the solid line represents the solution.
        The thick dashed line near the top of each profile shows the fit 
        by the self-similar closing regime (Eq. \ref{eq:Rasy}).
        Each sample is located in the phase diagram (a).
        All lengths are expressed in \micro\meter.
    }
    \label{fig:matching}
\end{figure}

\begin{figure}[p!]
    \centering
    \includegraphics[width=\linewidth]{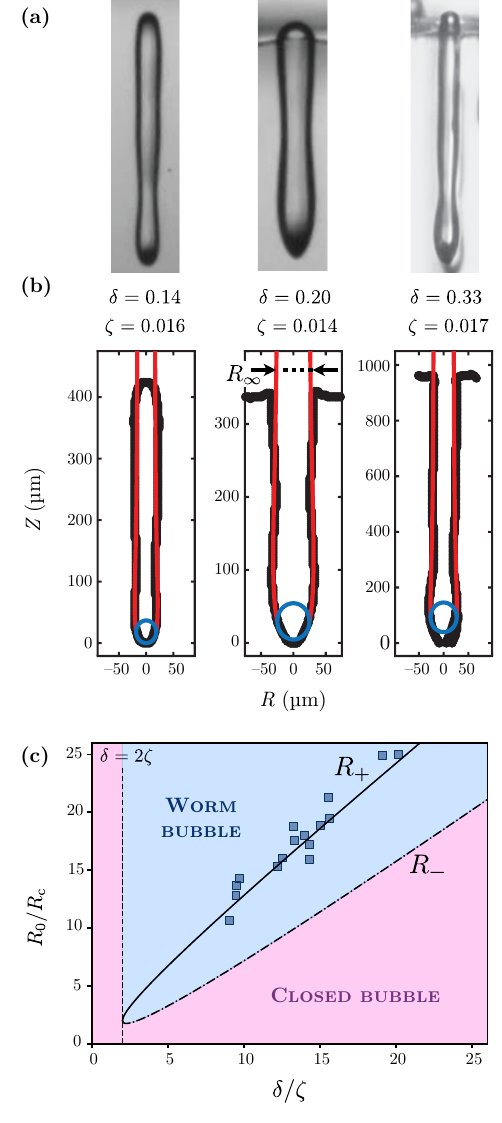}
    \caption{
        (a) Pictures of various worm bubbles
        with (b) their shapes matched (solution in red, initial condition in blue).
%         Left: $\delta = 0.14$ and $\zeta = 0.016$;
%         center: $\delta = 0.2$ and $\zeta = 0.014$;
%         right: $\delta = 0.33$ and $\zeta = 0.017$.
        (c) Phase diagram predicting whether the bubble will close (purple)
        or not (blue), depending on parameters $\delta$ and $\zeta$ and
        on the initial condition $R_0$.
        The equilibrium radius of worm bubbles $R_\infty$ must be equal to $R_+$ (Eq.~\ref{eq:poles}).
        Dark blue squares show the experimental measurements of $R_\infty$.
    }
    \label{fig:worms}
\end{figure}

\section*{Shape matching}
% \subsection*{Shape matching}
% \lipsum
\subsection*{Closed pores}
%%%%%%%%%%%%%%% Method
Our experiments give us access to pictures of frozen bubbles of which we know at
what rate they froze.
Unknown are the supersaturation \dc around the bubble when it appears
and the nucleation process.
In the following we match numerical solutions of the Frozen Bubble Equations
to the pores observed in our experiments. 
% The form given by Eq.~(\ref{eq:frozen_alt}) makes the matching more practical
% because $R_0$, $Z_0$ and $\Rp_0$ can be visualized on the pore profile.
For simplicity, we shall assume that nucleation and entrapment
leave the bottom of the bubbles spherical
-- that is, of uniform curvature --
up to the point ($Z_0$, $R_0$).
Therefore, in Eq.~(\ref{eq:frozen_alt}) the initial slope
can be expressed as $\Rp_0 = \left(R_0^2 - Z_0^2\right)/2 R_0 Z_0$.
We measure $Z_0$ and $R_0$ on the profiles and fit \Rc
by matching the overall shape of the bubble;
% We are left with only one fitting parameter, \Rc,
thence we compute $\zeta$, $\delta$ and \dc.

%%%%%%%%%%%%%%% 
After its shape is matched with a solution,
each pore may be placed in a phase diagram (Fig.~\ref{fig:matching}a).
This phase diagram confirms \emph{a posteriori}
the most important feature of the problem treated in this paper:
both $\delta$ and $\zeta$ are of order $1$,
therefore none of them may be neglected.
Both supersaturation and freezing must be taken into account to properly describe the pores.
This justifies the complexity of Eq.~(\ref{eq:frozen}).
It should be noted that removing the transient diffusion term
$\sqrt{\zeta/\pi Z}$ in Eq.~(\ref{eq:frozen}) prevents from 
matching the shapes.
Therefore, transient diffusion is important in the growth of the pore, 
contrary to recent assumptions~\cite{shao2023growth}.
Nevertheless, this two-dimensional phase diagram is insufficient to characterize the pore shapes;
two pores with similar $\delta$ and $\zeta$ but grown out of different initial conditions
may have significantly different shapes 
(\emph{e.g.} Fig.~\ref{fig:matching}b and f, or Fig.~\ref{fig:matching}c and g).

Most pores that we observe are well-matched from top to bottom.
Pores with a tail, due to nucleation between ice grains, 
are well-matched starting above the tail (Fig.~\ref{fig:matching}c).
This suggests that transient gas diffusion following the nucleation
is delayed until the bubble is free, and not stuck between ice grains.
Some large pores can only be matched partially from some height $Z_0$ to their tip
(Fig.~\ref{fig:matching}f-g).
In these cases, the bottom of the pore does not have a uniform curvature,
so our assumption that the bubble is left spherical by the entrapment probably
fails.
It is also likely that for very slow freezing,
wetting effects become noticeable and our assumption 
of a hemispherical bubble fails as well.

In addition to matching the full solution of the Frozen Bubble Equation,
we also fit the self-similar regime corresponding to the closing of the pore (Eq. \ref{eq:Rasy}).
In each of Figs. (\ref{fig:matching}b-g), a thick dashed line with a contrasting color shows the
good agreement between the various experimental profiles and the scaling law
$R\propto(Z_\mathrm{max}-Z)^{1/3}$.
% We obtain a similar agreement over the whole range of parameters investigated.
It should be noted that even in cases where we struggle to obtain a good match between the profile
and the full solution (like Fig. \ref{fig:matching}g), 
the self-similar regime near the top is recovered.

\subsection*{Measurements}
In the following, we consider pores frozen during a single experiment,
corresponding to Figure~(\ref{fig:shapes}c).
The large number of pores formed during the same experiment enables
to study their statistics \cite{supmat}.
These pores were frozen at a freezing rate $133~\micro\meter\per\second \pm 10~\micro\meter\per\second$;
% (average $\pm$ standard deviation);
in Figure~(\ref{fig:matching}a) they are situated in the main cloud of points.
Matching a solution of the Frozen Bubble Equation to each pore,
we can measure its nucleation radius;
$\Rc = 5.8\ \micro\meter \pm 0.7\ \micro\meter$.
From the definition of \Rc follows the supersaturation:
$\dc = 0.76 \pm 0.1\ \mathrm{g}/\mathrm{L}$,
which is more than 40-fold the initial concentration (Materials and Methods).

Shape matching enables to measure $R_0$ and $Z_0$ for each bubble;
we find $R_0 = 7.8\ \micro\meter \pm 1\ \micro\meter$ and
$Z_0 = 8.2\ \micro\meter \pm 1.4\ \micro\meter$.
These values close to the nucleation radius show that entrapment occurs
very shortly after nucleation.
The initial condition $R_0$ and $Z_0$ corresponds to some time $T_0$
after the bubble nucleation.
$T_0 = Z_0/v$ is the delay between the nucleation (when $R=0$) 
and the point at which $R=R_0$.
We measure $T_0 = 60\ \mathrm{ms}\ \pm 12\ \mathrm{ms}$.
% Interestingly, this shows that all the bubbles in this series
% have homogeneously nucleated ahead of the front.

In order to obtain a simple form for the Frozen Bubble Equation, we have assumed that
the bubble meets the freezing front with a right angle;
in reality, this angle is $\pi/2 + \varepsilon$ and we have neglected $\varepsilon$.
Shape matching enables to measure the initial slope $R_0^\prime$ that the bubble makes
with the freezing front, and that we assumed to be zero to simplify the description of diffusion.
We find that $R_0^\prime = -0.06 \pm 0.04$, corresponding to angle 
$\varepsilon_0 = -\arctan{R_0^\prime} = 3.4 \pm 2.3^\circ$,
which is indeed negligible.

Another remarkable result concerns the ratio $R_0/\Rc$,
which is the actual initial condition in the non-dimensional 
Eq.~(\ref{eq:frozen}).
Its value is $1.33 \pm 3\times 10^{-4}$, which is an extremely narrow range.
This is likely due to the conditions in which the bubble nucleates at the freezing front. 
A very narrow range of values suggests that the nucleation crevices \cite{atchley1989crevice}
have very similar sizes and shapes in the given range of freezing rate.
More analysis is required to describe these results quantitatively.
% Although we are not able to explain this value,
% it is likely linked to the entrapment mechanism.

%%%% Worm bubbles
\subsection*{Worm bubbles}

Shape matching can also be applied to worm bubbles.
In Figure \ref{fig:worms}, we use this technique to measure 
$\delta$, $\zeta$ and \Rc for various worm bubbles,
as well as their equilibrium radius $R_\infty$ (Fig.~\ref{fig:worms}b), during or after their growth.
For the latter we neglect the closed tips that occur due to the above-mentioned fluctuations.
According to Eq. \ref{eq:separatrix} and \ref{eq:poles},
worm bubbles occur when two conditions are fulfilled: $\delta > 2\zeta$ and $R_0/\Rc > R_-$.
In this case, the stable radius of the worm bubble is equal to $R_+$.
Placing our measurements on a phase diagram for the bifurcation (Fig. \ref{fig:worms}c)
confirms that our model accurately predicts when worm bubbles appear as well as 
their equilibrium radius.

%%%%%%%%%%%%%%%%%%%%%%%%%%%%%%%%%%%%%%%%%%%%%%%%%%%%%%%%%%%%%%%%%%%%%%%%%%%%%%%%
%%%%%%%%%%%%%%%%%%%%%% CONCLUSION %%%%%%%%%%%%%%%%%%%%%%%%%%%%%%%%%%%%%%%%%%%%%%
%%%%%%%%%%%%%%%%%%%%%%%%%%%%%%%%%%%%%%%%%%%%%%%%%%%%%%%%%%%%%%%%%%%%%%%%%%%%%%%%
% \clearpage
\section*{Conclusion}

The shape of gas bubbles trapped in ice results of simultaneous 
freezing and growth by gas diffusion.
Heat transfer and capillarity set the shape of the bubble
and of the freezing front.
Depending on the bubble size relative to the nucleation radius \Rc, 
diffusion makes it grow or shrink.

We have demonstrated that the shapes of bubbles trapped in ice, 
although extensively diverse,
can be accurately described by a single non-linear ordinary differential equation,
the Frozen Bubble Equation (Eq.~\ref{eq:frozen}).
The non-dimensional parameters $\delta$ and $\zeta$,
respectively representing the supersaturation and the freezing velocity,
suffice to describe the growth of a bubble from a given initial condition $R_0 = R(Z_0)$.
The asymptotic regimes explain why the tip of the pores is so
characteristically rounded -- it follows a power law (Eq.~\ref{eq:Rasy}) --
and why the quickly frozen bubbles tend to be slightly elongated
-- the limit shape absent diffusion is a cone. 
Matching a solution of the Frozen Bubble Equation to the shape of a real pore enables to measure
the supersaturation and the nucleation radius at which the pore appeared.
We have shown that this is at least possible for freezing rates in the range 
12~\micro\meter\per\second\ to 263~\micro\meter\per\second.

The mathematical analysis reveals a bifurcation that explains how worm bubbles,
these cylindrical pores of potentially several centimeters, are formed.
To the best of our knowledge,
this is the first analytical model to make such predictions.
It yields the conditions for worm bubbles to appear as well as their equilibrium radius.
Both are confirmed by our experiments.
In further work, it would be interesting to validate the model
against measurements made using different gases, with different solubility, such as CO$_2$ (higher solubility) or argon (lower solubility),
and in a wider range of freezing rates.

Our model is written in the most parsimonious fashion;
it describes the pores well with as few parameters and mechanisms as possible.
It could be extended to take into account the impact of neighbor bubbles on the
concentration field~\cite{yoshimura2008}.
Mathematical analysis could also be extended to the case in which the freezing rate
is not constant but decreases as the freezing front moves away from the thermostat~\cite{thievenaz2019a}.
Our work is applicable to measuring the freezing history of porous ice.
It could also be used to help the design of porous freeze-cast materials.

\section*{Materials and Methods}
\subsection*{Solution}
We use deionized water left in contact with air for a few days 
so that gases dissolve in it.
Before the experiment, the concentration of oxygen in the water
was measured at 
$c_{O_2} = 6.7\: \mathrm{mg}/\mathrm{L}$ at $21 ^\circ\mathrm{C}$.
Using known correlations of the solubility and Henry constant 
with the temperature~\cite{lide2012}, 
we obtain the concentration of nitrogen $c_{N_2} = 11.6\: \mathrm{mg}/\mathrm{L}$,
and the concentration of air as a mixture of both gases
$\cnot = 18.3\: \mathrm{mg}/\mathrm{L}$.
% Similarly, the solubility of oxygen in water at $0^\circ\mathrm{C}$ is
We do not take into account further differences between oxygen and nitrogen.

\subsection*{Experiments}
Our experiments consist in freezing water contained in a Hele-Shaw cell
(Fig.~\ref{fig:shapes}a).
We have used two seperate experimental setups.

In the main setup, we used glass capillary tubes of rectangular section (Vitrocom), 
with inner dimensions
6~mm by 300 \micro\meter\ for the bigger cell,
2~mm by 100 \micro\meter\ for the smaller cell.
No significant effect on the bubble shape of using one or the other cell was found.
The freezing of the water column is recorded with a DSLR camera (Nikon D5600) mounted with
a macro lens (Nikon Micro-Nikkor AI-s 200mm f/4) and a microscope lens (Mitutoyo).
The whole set-up is backlit by a light panel (Phlox).
The capillary is first filled up with water and then carefully 
brought in contact with a thermostat, 
whose temperature is kept constant at $-25^\circ\mathrm{C}$ throughout the experiment.
The thermostat is a hollow copper plate through which cold oil 
is pumped from a refrigerating bath (Julabo Corio 1000F).
The local rate of freezing is set by the rate at which the latent heat released 
at the ice-water interface diffuses through the ice to the thermostat~\cite{thievenaz2019a}.
Therefore, for a given heat flux absorbed by the thermostat,
$v$ decreases with the distance to the thermostat and with the section of the Hele-Shaw cell.
We measure the local rate of freezing $v$ on each video, next to each bubble.
For all the experiments that we discuss in the present paper, $v$ remains constant 
(within a few \micro\meter\per\second) during the formation of each bubble; 
it ranges from 12 to 263~\micro\meter\per\second.

The second experimental setup, used to study the formation of worm bubbles,
follows a similar design and protocol (see Supplementary Material).
In short, room temperature water is deposited between two acrylic plates 
(spaced 1 mm apart) 
on a cooled and frosted substrate, which is mounted on top of a freezing stage (BFS-40 MPA, Physitemp). 
The temperature of the substrate at the base of the deposited water is measured by a thermocouple.
By varying the temperature of the substrate we control the range of freezing velocities,
here limiting ourselves to a substrate temperature of $-7.5^\circ\mathrm{C}$,
yielding velocities between 4~\micro\meter\per\second\ to 17~\micro\meter\per\second,
ensuring the formation of worm bubbles. 
The freezing process is recorded in side-view using a camera (Nikon D850)
connected to a long working distance lens (Thorlabs, MVL12X12Z). 
The sample is illuminated with a diffused cold-LED to avoid local heating.

\subsection*{Numerical resolution}
The Frozen Bubble Equation was integrated using a fourth-order Runge-Kutta scheme in
a custom Python routine.

\subsection*{Compressibility effects}
In the derivation we have assumed that air has constant density.
Compressibility can be taken into account when we develop the time derivative of the mass
in Eq.~(\ref{eq:mass_balance}), 
by introducing the isothermal compressibility 
$\chi = \rho^{-1}\left(\partial \rho/\partial P\right)$.
The calculation yields a slightly different version of Eq.~(\ref{eq:frozen}):
\begin{equation}
    \frac{\mathrm d R}{\mathrm d Z} =
%     \delta \frac{\left( 1 - \frac{1}{R} \right)}{\left(1 - \frac{R_\chi}{\Rc}\cdot\frac{1}{R}\right)}
    \delta \left(\frac{ R-1 }{R - \frac{R_\chi}{\Rc}}\right)
    \left( \frac{1}{R} + \sqrt\frac{\zeta}{\pi Z}\right)
    - \frac{1}{2},
    \label{eq:frozen_chi}
\end{equation}
where $R_\chi = 2\gamma\chi/3$ is the length scale associated to compressibility.
The first term of Eq.~(\ref{eq:frozen_chi}) resembles Eq. 6 of reference \cite{bari1974},
which follows from a different derivation that starts from the equation of state of the gas.
In the limit $R_\chi \ll \Rc$ Eq.~(\ref{eq:frozen}) is recovered.
For air $\chi \simeq 10^{-5}\ \mathrm{Pa}^{-1}$, so $R_\chi \simeq 0.5\ \micro\meter$.
In our experiments \Rc is at least ten times larger,
therefore compressibility is negligible compared to dissolution.

\section*{Acknowledgments}
This material is based upon work supported by the National Science Foundation 
under NSF Faculty Early Career Development (CAREER) Program Award CBET No. 1944844,
and by a UCSB Senate Faculy Grant.
VT and AS thank Sylvain Deville and C\'ecile Monteux for fruitful discussions
at the early stages of the project, and Sylvain Deville especially for reviewing
the first version of the manuscript.
VT warmly thanks Laurent Duchemin for insightful discussions about the mathematical methods.
JM and DL thank the Balzan Foundation for support.

\subsection*{Author contributions}
This article results of the merger of two distinct projects conducted by
VT and AS on one side and JM and DL on the other side.
VT constructed the model, conducted the mathematical analysis, 
and performed the experiments with the first experimental setup.
JM conducted the experiments with the second experimental setup,
especially concerning the worm bubbles.
VT and AS wrote the manuscript, JM and DL reviewed it.

\bibliography{frozen_bubbles}

\clearpage
\appendix
\section{Supplementary Information}
\subsection{Videos}

Figure 1 is supplemented with videos showing the formation of each pore,
labeled with the number of the subfigure, from 1b to 1i.
These videos are all played in real time. \\

Figure 6 is also supplemented with videos showing the formation of worm bubbles.
\begin{itemize}
\item Video 1: Typical experimental footage of the freezing process showcasing 
    the growth process of worm bubbles.
\item Video 2: Growth of the worm bubble corresponding to Figure~6a of the main text (second panel).
\item Video 3: Growth of the worm bubble corresponding to Figure~6a of the main text (third panel).
\end{itemize}

\subsection{Experimental setup to study worm bubbles}

\begin{figure*}
  \centering
  \includegraphics[width=.95\textwidth]{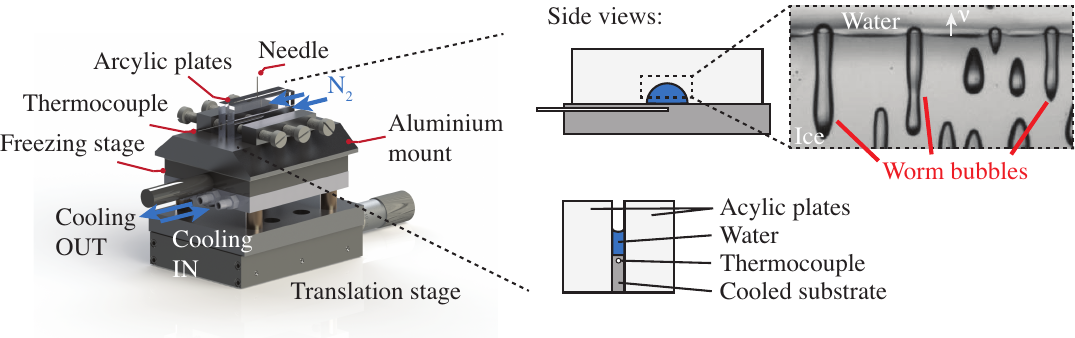}
  \caption{Schematic of the experimental Hele-Shaw set-up (left) including side-view sketches and an experimental snapshot that depicts the growth of worm bubbles in ice (right). During this process the solidification front advances upwards with velocity $v$. }
\label{fig:1}
\end{figure*}

To study the formation and growth of worm bubbles we make use of a different setup 
with which slower freezing rates can be reached in a controlled way.
The aim of the experimental set-up is to freeze a sessile water drop on a cold substrate.  
During the freezing process, bubbles will naturally nucleate and grow near the advancing solidification front, 
eventually leading to the formation of worm bubbles.
To avoid lensing effects we are interested in freezing only a thin slice of purified water (Milli-Q).
In order to achieve this, an aluminium mount is placed on top of a freezing stage (BFS-40 MPA, Physitemp) 
that allows for two acrylic plates to be pressed against a thin metal strip (see Figure\,\ref{fig:1}).
The gap between the plates is 1 \milli\meter and the temperature of the substrate 
close to the base of the deposited water is measured by a thermocouple 
that is placed inside a groove at the side of the metal strip.
We make sure that the desired bottom temperature, here \unit{-7.5}\celsius, 
has been reached well within $\pm\,\unit{0.1}\kelvin$ for several minutes before starting the experiment.
A needle (Nordson) and a syringe pump (PHD 2000 Infusion, Havard Apparatus) 
are used to deposit amounts of water of equal volume ($V_\mathrm{d} = \unit{25}\micro\liter$) between the plates, 
resulting in a 'drop' with a typical height of roughly $\unit{3}\milli\meter$. 
To guarantee freezing as soon as the (room temperature) water touches the substrate and to avoid supercooling, we only deposit the water once ice crystals have formed on top of the thin metal strip.  A gentle flow of nitrogen along the outsides of the plates prevents fog and frost formation that otherwise would obscure the view.
The drop is illuminated with a diffused cold-LED to avoid local heating.
The freezing process is recorded in side-view using a camera (Nikon D850) connected to a long working distance lens (Thorlabs, MVL12X12Z). 
A typical experimental snapshot is shown in figure\,\ref{fig:1} that depicts the growth of worm bubbles.

\subsection{The limit $R \to 0$}
\begin{figure*}[h]
    \centering
    \includegraphics[width=\textwidth]{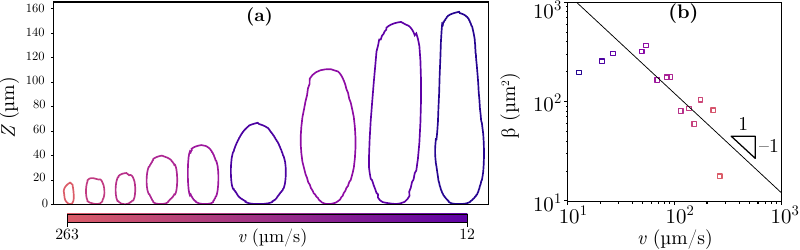}
    \caption{(a) Samples of pore profiles to highlight the asymptotic regime when $R \to 0$.
%     (b) Rescaling (Figure 3b of the main text).
    (b) Evolution of parameter $\beta$ with the freezing rate $v$.}
    \label{fig:SMclosing}
\end{figure*}

Figure \ref{fig:SMclosing}a shows the profiles described in Figure \ref{fig:asympt}.
Figure \ref{fig:SMclosing}b shows values of $\beta$ obtained through the rescaling.

\subsection{Measurements on the pores of Figure 1c}
\begin{figure*}[h]
    \centering
    \includegraphics[width=\textwidth]{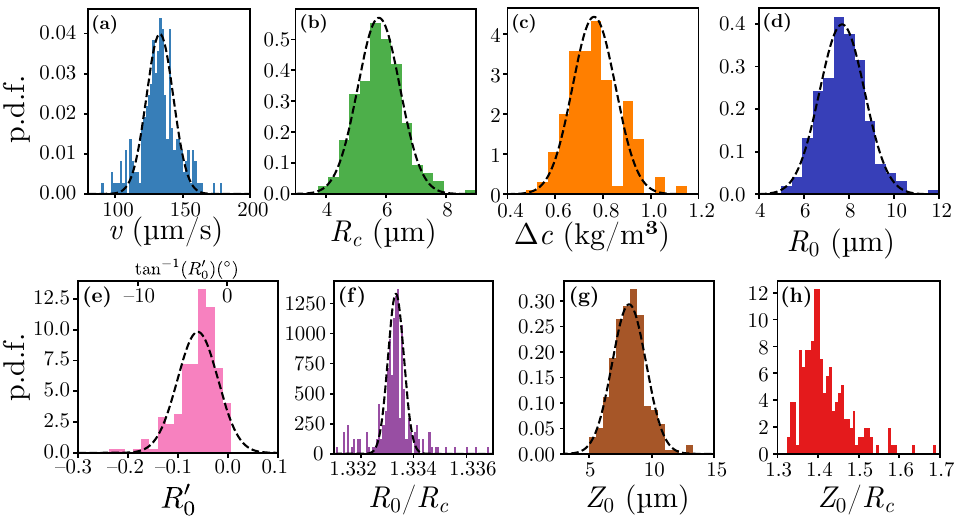}
    \caption{Statistics of the measurements made by shape matching on the pores of Figure 1c.
    Dashed lines are best Gaussian fits to guide the eye.}
    \label{fig:SMmeasurements}
\end{figure*}

In addition to the discussion of the dimensions of the pores of Figure \ref{fig:shapes}c
we provide the statistics of measurements performed thereon, in Figure \ref{fig:SMmeasurements}.

\end{document}